\documentclass[useAMS,usenatbib]{mnras}
\usepackage[usenames,dvips]{color}
\usepackage{amsmath}
\usepackage{amssymb}
\usepackage{graphicx}
\usepackage{epsfig}
\usepackage{hyperref}

\usepackage[dvipsnames]{xcolor}

\usepackage{url}
\usepackage{indentfirst}
\usepackage{color}
\usepackage[normalem]{ulem}
\usepackage[left]{lineno}

\newcommand{\NH}{V1674~Her}

\title[High-energy orbital timing features of V1674 Her]
{Investigation of a likely orbital periodicity of Nova Hercules 2021 in X-rays and $\gamma$-rays}       
\author[L. C.-C. Lin, Jhih-Ling Fan, Chin-Ping Hu, Jumpei Takata and Kwan-Lok Li]{Lupin Chun-Che Lin$^1$\thanks{E-mail:lupin@phys.ncku.edu.tw}, Jhih-Ling Fan${}^1$, Chin-Ping Hu${}^{2}$, Jumpei Takata$^3$ and Kwan-Lok Li$^1$\thanks{E-mail:lilirayhk@phys.ncku.edu.tw}\\
$^1$Department of Physics, National Cheng Kung University, Tainan 701401, Taiwan\\
$^2$Department of Physics, National Changhua University of Education, Changhua 50007, Taiwan\\
$^3$Department of Astronomy, School of Physics, Huazhong University of Science and Technology, Wuhan 430074, China\\
}

\begin{document}
\date{August 2022}
\pagerange{\pageref{firstpage}--\pageref{lastpage}}
\pubyear{???}
\maketitle
\label{firstpage}


\begin{abstract}
We report a detection of a $\sim$0.153\,days period in the classical nova \NH\ using the \emph{NICER} observations taken within a month since the outburst (i.e., $\sim$MJD~59405). The X-ray period is consistent with the orbital period previously found in the optical band, strongly suggesting the \emph{NICER} signal as the X-ray orbital periodicity of the system.
A seemingly double-humped profile was obtained by folding the detrended X-ray light curve with the period after removing the rotational X-ray pulsations of the nova.
The profile may be caused by occultation by the companion or the accretion disk, possibly indicating a high inclination of the system.
The $\gamma$-ray emission of \NH\ with a significance level $\gtrsim 5\sigma$ was detected by $Fermi$-LAT close to its optical peak and the emission faded away within 1 day, which is the shortest duration known for a $\gamma$-ray nova.
Folded on 0.153\,days, a marginal $\gamma$-ray variability can be also seen in the LAT light curve, but without the double-hump feature observed in X-rays. 
If the $\gamma$-ray modulation is real, its origin is probably different from that observed in the X-ray and optical bands.
\end{abstract}


\section{Introduction}
\label{sec:intro}

Nova Herculis 2021 (hereafter \NH) was discovered as TCP~J18573095+1653396 by Seiji Ueda on 2021 Jun. 12.5484 UT, with an apparent magnitude 8.4. A 16.62-mag pre-discovery detection from the All-Sky Automated Survey for Supernovae (ASAS-SN) showed that the transient actually started at least 8.4 hours before the discovery (2021 Jun. 12.1903 UT; \citealt{Aydi2021}), and it was soon identified as a classical nova with optical spectroscopy \citep{MVD2021}.
\NH\ was a fast nova, of which the magnitude dropped by 2 magnitude from the peak of $V\approx6$ mag in one day \citep{MVD2021,Quimby2021}, and could be the ``fastest nova on record'' \citep{Wagner2021}.
\NH~can be seen in multiple wavelengths from GHz radio to GeV $\gamma$-rays \citep{Sokolovsky2021,Wagner2021,Woodward2021,Kuin2021,Li2021}. The X-ray telescope (XRT) of the \emph{Neil Gehrels Swift Observatory} (\emph{Swift}) began to detect \NH\ on 2021 Jun. 14.41. 18 days later, it became a supersoft X-ray source (SSS; \citealt{Page2021}), and the XRT spectrum can be well fitted by a blackbody model \citep{Drake2021} with temperatures of $\lesssim100$~eV.
In addition, \emph{Neutron Star Interior Composition Explorer} (\emph{NICER}) and \emph{Chandra} Low-Energy Transmission Grating (LETG) confirmed that \NH\ had entered the supersoft X-ray phase \citep{Drake2021,Orio2022}.

One of the most interesting features of \NH\ is its timing properties.
A spin period of 501.42\,s (i.e., 8.357\,min) can be detected using the archival Zwicky Transient Facility (ZTF; \citealt{Bellm2019}) data collected in the pre-outburst stage \citep{Mroz2021}.
A similar signal at 501.52(2)\,s (i.e., 8.3586(3)\,min) can also be detected with the time-series photometry observed by the globally distributed telescopes of the Center for Backyard Astrophysics (CBA) during the outburst (i.e., 2021 Jul.-- 2021 Aug.; \citealt{Patterson2021}). 
In addition to the optical band, $\approx500$-s X-ray pulsations were detected by \emph{Chandra} \citep{Maccarone2021,Drake2021}, \emph{Swift} \citep{Page2021}, and \emph{NICER} \citep{Orio2022}, suggesting that \NH\ is an intermediate polar (IP), which hosts a magnetized white dwarf (WD) with a surface magnetic field strength of several $10^5$\,G \citep{Orio2022}. No obvious spin-down of \NH\ can be found \citep{Orio2022}. Besides the spin signal, a double-humped orbital profile with a period of 0.15302(2)\,d was detected in the optical band \citep{Patterson2021}. However, the community paid less attention to the multi-wavelength analysis on the orbital signal, though the phenomenon is also rare for classical novae. Therefore, in this letter, we present the investigation of the orbital period in the X-ray band using \emph{NICER} data and the variation of the $\gamma$-ray counterpart for \NH. 
The possible scenarios of the emitting nature are also discussed in the final section.

 
\section{Data Reduction and Analysis}
\label{sec:observations}

In this letter, we concentrate on the investigation of \NH~in the X-ray band and its $\gamma$-ray counterpart. 
To accomplish our study, we used the X-ray archive of \emph{NICER} observations obtained from NASA’s High Energy Astrophysics Science Archive Research Center (HEASARC) and performed the related analyses using the HEASoft package \citep{HEAsoft2014}.
In the $\gamma$-ray band, we analyzed the  Pass 8 (P8R3) data collected by the Large Area Telescope (LAT). 
To study the high-energy emission properties of \NH, we extracted the $\gamma$-ray source photons centered at R.A.=$18^{\rm h}57^{\rm m}30\fs{98}$, decl.=$16^{\circ}53'39\farcs5$ (J2000) according to the ASAS-SN sky patrol \citep{Aydi2021}.
For the timing analysis, we corrected the instrumental time to the barycentric dynamical time (TDB) using the JPL DE405 solar system ephemeris at the aforementioned source position.

\subsection{{\sl \textbf{NICER}}}
\label{ssec:NICER}
\emph{NICER} monitored \NH~with a series of observations using the X-ray timing instrument (XTI) from 2021 Jul. 10 to Aug. 31.
Before Aug. 12, \emph{NICER} observed the target everyday with $\sim$0.5--16\,ks exposures except for Jul. 31 and Aug. 8.
From Aug. 20 to 31,  \emph{NICER} did not have observations on Aug. 23--24 and Aug. 27, and the exposures ranged from 0.5 to 2.7\,ks.
Although \emph{NICER} does not have an imaging capability, it is enabled a precise temporal resolution ($< 300$\,ns) and a high sensitivity in 0.2--12\,keV to investigate the periodic signal embedded in \NH~\citep{Okajima2016}.
   
Photons collected from all the 52 \emph{NICER} focal plane modules (FPMs) were all kept for timing analysis, but we only extracted them in the energy range of 0.25--12\,keV to avoid the significant noise contamination below this energy range.
We performed the reduction of all the \emph{NICER} observations using HEASoft (v.6.29) and the barycentric time correction to the photon arrival times using the \texttt{barycorr} task.
We note that the target's brightness significantly became weaker since 2021 Jul. 31 (i.e., MJD 59425) so we also re-binned the data to trace the variability of the source.   

\subsection{{\sl \textbf{Fermi}}}
\label{ssec:Fermi}
In order to investigate the $\gamma$-ray counterpart of \NH\ reported by \cite{Li2021}, we re-analysed the {\it Fermi} Large Area Telescope (LAT) Pass 8 (P8R3) data in the energy range of 0.1--300 GeV.
From the Fermi Science Support Center (FSSC), we downloaded the data obtained within an interval from 2021 June 5 (MJD~59370; one week before the nova discovery) to 2021 July 12 (MJD~59407; one month after the discovery). 
Source photons were extracted using a circular region of interest (ROI) of radius $10^{\circ}$ centered at the nova optical position. All the analysis processes were done by the \emph{Fermi} Science tools (v11r5p3) .

We selected photons that belong to the class for point source or Galactic diffuse analysis (i.e., event class 128) and were collected in the front- and back-sections of the tracker (i.e., event type 3) in our analysis.
The instrument response function applied for the selected event type is ``P8R3\_SOURCE\_V2'', which do not have an anisotropic component of the residual background mainly caused by electrons in comparison to the earlier functions.
We only used photons obtained in the good-time-intervals (i.e., DATA\_QUAL $>$ 0) of the spacecraft and removed events with zenith angles larger than $90^{\circ}$ to avoid contamination from Earth's albedo $\gamma$-rays. 
The binned likelihood analysis method provided by the \emph{Fermi} science team was adopted with an emission model, which contains all the 4FGL-DR3 $\gamma$-ray sources \citep{Abdollahi2022} within 15 degrees from the nova and two diffuse background components (i.e., \texttt{iso\_P8R3\_SOURCE\_V3\_v1} and \texttt{gll\_iem\_v07}), generated by the FSSC user-contributed Python script \texttt{make4FGLxml.py}.
Since the \NH\ was brightest $\gamma$-ray source in the field\footnote{The $\gamma$-ray emission reached $\sim10^{-10}$~erg\,cm$^{-2}$\,s$^{-1}$ at peak \citep{Li2021}, and there is no cataloged $\gamma$-ray source of $>10^{-10}$~erg\,cm$^{-2}$\,s$^{-1}$ within 5 degrees from the nova according to 4FGL-DR3 \citep{Abdollahi2022}.}, all the spectral parameters of the 4FGL-DR3 sources in the emission model were fixed for simplicity.

\begin{figure}
\centering
\includegraphics[width=8.2cm,height=5.6cm]{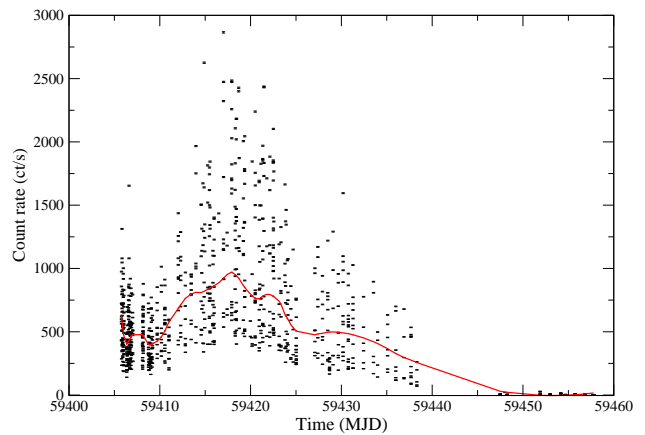}
\caption{Light curve of the \NH~obtained from the \emph{NICER} observations binned with 100\,s. Uncertainties are too small to be clearly labelled in the figure. The red curve demonstrates the evolutionary trend of light curve.} 
\label{LC-bin100}
\end{figure}

\section{Results}
\label{sec:result}

\subsection{{\sl \textbf{X-ray timing signals}}}
\label{ssec:XPeriodicity}

The spin period of \NH~was detected in the optical band \citep{Bellm2019} and can also be confirmed in the X-ray band via different observations including \emph{Swift} \citep{Page2022}, \emph{Chandra} \citep{Maccarone2021,Drake2021} and \emph{NICER} \citep{Pei2021,Orio2022} even though the source returned to a quiescence stage in 2022.
In comparison to the investigation of the evolution of the spin period, the candidate signal of the orbital period was only reported in the optical band \citep{Patterson2021}.
We check the orbital periodic signal using the \emph{NICER} observations to monitor the source for almost two months soon after the  outburst in 2022 Jun.
We notice that both \emph{Swift} and \emph{NICER} provide the long-term observation to our target; however, the larger effective area and a better sensitivity of \emph{NICER} are helpful in detecting the orbital period that is superimposed with the spin signal.

We used 100\,s to bin the data summarized in Section~\ref{ssec:NICER} as shown in Fig.~\ref{LC-bin100}, and the red curve was determined by the local regression of nearby 200 data points using weighted linear least squares and a 2nd order polynomial model.  
In the preliminary test, we considered to check the orbital period in a short time interval with the most dense data points (i.e., 2021 Jul. 10--13) by the Lomb-Scargle periodogram (LSP; \citealt{Lomb76,Scargle82}).
We can obtain two candidate signals with 99.9\% white noise significance level \citep{HB86} at the period of 0.1536(6)\,d and 0.1110(3)\,d. 
For the following time intervals, the data points are relatively sparse and the average count rate of each observation has a significant flux variation so the LSP is much more noisy.
We speculated to remove the global trend of the variation as the red curve shown in Fig.~\ref{LC-bin100}, and the LSP obtained from the residual light curve for 2021 Jul. 10--13 enhances the Lomb-Scargle power of two major signals; however, the signal of the frequency between 6--7\,$d^{-1}$ can still be detected to correspond to 0.1528(3)\,d with 90\% white noise significance level using 2021 Jul. 15--25 observations, but another one became insignificant.

\begin{figure}
\centering
\includegraphics[width=9.5cm,height=6.5cm]{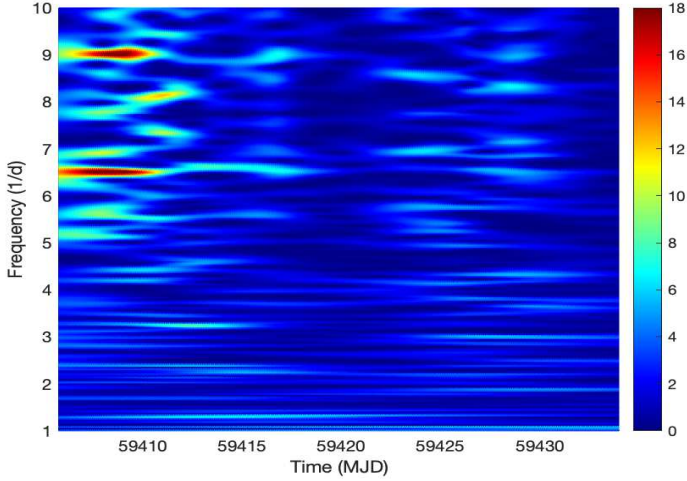}
\caption{WWZ periodogram of \NH. The periodogram was generated with WWZ using the detrending light curve. The colour of contour denotes the strength of the power given by WWZ.} 
\label{WWZ}
\end{figure}

Even though we remove the global trend of the X-ray variation for \NH, the periodic signal is still difficult to be clearly resolved with the LSP by the observations after 2021 Jul. 25.
The major reason is that we have fewer data points within a much longer time span after Jul. 25; for instance, we have similar data points within 2021 Jul. 10--13 and Jul. 15--25, and it leads to a significantly drop on the detected power.
To investigate the variation of the signal of $\sim$0.153\,d, we selected the kernel parameter to characterize a small Gaussian envelope ($c=10^{-4}$; \citealt{Foster96}) to perform the weighted wavelet z-transform, which is a dynamical timing analytical scheme as described in \citet{Lin2015}.
Fig.~\ref{WWZ} demonstrates the result obtained from the WWZ periodogram, and we can see a signal with a frequency between 6--7\,d$^{-1}$ across the entire investigation corresponding to the period of $\sim$0.153\,d.
The signal shows a tendency to slightly increase its frequency since MJD~59410 (i.e., i.e., 2021 Jul. 15), and the power of the signal significantly decreased since MJD~59420, which are consistent to the results obtained from the LSP.

\begin{figure}
\centering
\includegraphics[width=8.2cm,height=5.6cm]{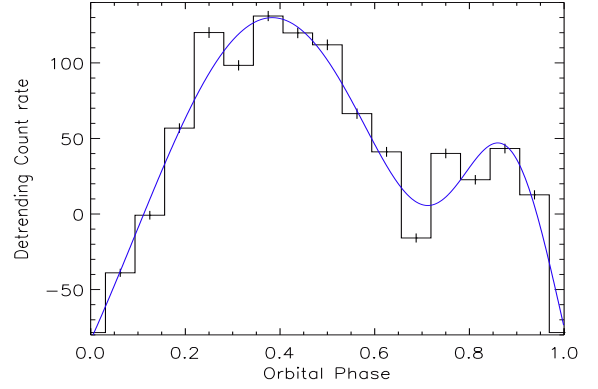}
\caption{The detrending X-ray folded light curve generated from the off-pulsed emission of \NH. The data were gained from \emph{NICER} light curve binned with 50\,s in 2021 Jul. 10--25 after removing the global evolutionary trend and the pulsed component. The value of each bin shown in the folded light curve was obtained with the mean value, and we also label the uncertainty of each bin. The blue curve presents the best-fit to a multi-Gaussian model. The epoch zero was arbitrarily determined to clearly demonstrate a double-humped feature.             
} 
\label{FLC}
\end{figure}

We tried to fold the detrended X-ray light curve with 0.153\,d. 
No significant orbital modulation feature can be seen, and the major variability feature could originate from the scattered data points due to the spin pulsation (ref. Fig.~\ref{LC-bin100}).
We therefore removed the pulsed component according to the timing solutions reported in the Table~6 of \citet{Orio2022} and regenerated the light curve using a time bin of 50\,s with the photons constrained in the off-pulsed phase.
Following the aforementioned procedures to detrend the light curve, we folded the light curve with the same signal only using the data collected before 2021 Jul. 25, during which the signal is more prominent (ref. Fig.~\ref{WWZ}).
The modulation feature is still not clear enough to be directly determined by the best-fit to a sinusoidal or a Gaussian model, but the distribution of data points in the folded light curve has a marginally visible trend.
Such a trend can be roughly resolved by an orbital profile composed of the mean value of data points in each bin of the folded light curve shown in Fig.~\ref{FLC}.
Please note that the small error bar of each bin only labels the uncertainty of the accumulated count rate, and it does not represent the dispersion scale of data points in one bin.
Here we also provided a best-fit multi-Gaussian model to the profile, which has a double-humped structure similar to the optical detection. 

\subsection{{\sl \textbf{$\gamma$-ray variations/counterpart}}}
\label{ssec:Gsource}

\begin{figure}
\centering
\includegraphics[width=8.2cm,height=5.6cm]{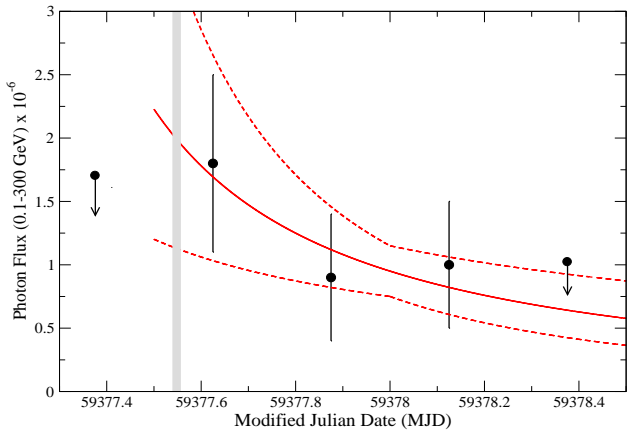}
\caption{The \emph{Fermi}-LAT light curve of \NH\ during the 0.75-day detection. The red line indicates the decreasing trend of the emission while the dashed lines presents the uncertainty interval. The gray region labels the discovery epoch of \NH\ at MJD 59377.5484.} 
\label{fig:fermi_lc}
\end{figure}

To check the $\gamma$-ray active period of \NH, a preliminary daily $\gamma$-ray light curve was first extracted assuming a simple power-law with a fixed photon index $\Gamma_\gamma=-2.2$ (which is close to that of other known $\gamma$-ray novae, e.g., \citealt{Abdo2010,Ackermann2014,Li2017}). 
The $\gamma$-ray emission only showed up on June 12 and 13 (MJD 59377 and 59378) with $\rm{TS}>4$ ($\rm{TS}=21.7$ and $7.2$, respectively; $\sqrt{\rm{TS}}\approx$ the detection significance in the unit of $\sigma$). 
We broke the 2-day interval into eight 0.25-day bins for a further investigation and found that the $\gamma$-ray emission was only detected (i.e., $\rm{TS}>4$) in 0.75 days from June 12.5 (MJD 59377.5) to June 13.25 (MJD 59378.25). 
A stack analysis\footnote{The \texttt{iso\_P8R3\_SOURCE\_V3\_v1} component was fixed to 1 in this analysis. Otherwise, it goes to zero.} with the 0.75-day data gives $\Gamma_\gamma=-2.3\pm0.2$ and $F_{\rm ph,0.1-300GeV}=(1.3\pm0.4)\times10^{-6}$~photon\,cm$^{-2}$\,s$^{-1}$ with $\rm{TS}=34.1$. 
Using the improved spectral emission model with the photon index fixed to the best-fit value, we updated the three 0.75-day detections, and the corresponding values are $F_{\rm ph,0.1-300GeV}=(1.8\pm0.7)\times10^{-6}$~photon\,cm$^{-2}$\,s$^{-1}$ ($\rm{TS}=17.0$), $(0.9\pm0.5)\times10^{-6}$~photon\,cm$^{-2}$\,s$^{-1}$ ($\rm{TS}=6.2$), and $(1.0\pm0.5)\times10^{-6}$~photon\,cm$^{-2}$\,s$^{-1}$ ($\rm{TS}=8.5$), respectively. 
Figure~\ref{fig:fermi_lc} shows the $\gamma$-ray light curve with a best-fit power-law of $F = (9.5\pm2.0)\times(t-t_0)^{-1.23\pm0.55}\times10^{-7}$~photon\,cm$^{-2}$\,s$^{-1}$ (where $t$ is the time in MJD and $t_0$ was defined as MJD 59377) to describe a rapid decline.

We also performed phase-resolved analyses for the spin (501.42 seconds; \citealt{Bellm2019}) and orbital (0.153~days; this work) periods using the 0.75 days of LAT data. 
While phase-resolved analyses are generally difficult for novae given the short time interval, and hence, the poor photon statistic (note that the recent $\approx$544\,s detection from ASASSN-16ma is the only candidate so far; \citealt{Li2022}), a $\gamma$-ray nova with known orbital/spin periods like \NH\ is extremely rare and this motivated us to work on $\gamma$-ray timing analysis. 
We employed the FSSC user-contributed Perl code, \texttt{like\_bphase}, which is a straightforward script to compute the exposure-corrected phased light curves for given LAT observations and periods. 
Besides, we manually computed 95\% upper limits for the light curve bins with $\rm{TS}<2$.
The $\gamma$-ray emission of \NH\ was decreasing rapidly in the 0.75-day interval, and the phased light curves can also be affected by the long-term trend. 
To eliminate the effect, we detrend the phased light curves from \texttt{like\_bphase} using the aforementioned power-law light curve model as well as the LAT exposures in the interval computed by \texttt{gtexposure}. 
Figure \ref{fig:period_lc} shows the final results with the phase zero determined at MJD 59377.5. 
In the orbital one, it seems that the $\gamma$-ray emission was a bit brighter in the phase interval of 0--0.4, although the variation is not significant after counting error bars.

\section{Discussions}
\label{sec:discussion}


\begin{figure}
\centering
\includegraphics[width=8.2cm,height=5.6cm]{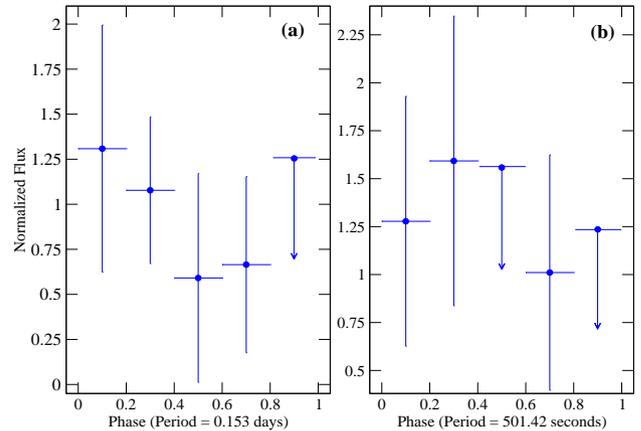}
\caption{The \emph{Fermi}-LAT phased curve of \NH\ folded with (a) 0.153 days and (b) 501.42 seconds. The flux has been normalized using the trend of the flux decline and the \emph{Fermi}-LAT exposures. Phase zero is determined at MJD 59377.5.} 
\label{fig:period_lc}
\end{figure}


Some SSSs/Novae also have orbital periods detected in the X-ray band (e.g., 0.0593\,d of Nova Muscae 1983, a.k.a. GQ Mus; \citealt{Kahabka96}, 0.1238\,d of V5116 Sgr; \citealt{Sala2017}, 0.1719\,d of 1E0035.4--7230; \citealt{Kahabka96}, 0.279\,d of V4743 Sgr; \citealt{Leibowitz2006}, 0.295\,d of V959 Mon; \citealt{Peretz2016}, and 1.77\,d of HV Ceti; \citealt{Beardmore2012}).
These detections can have different origins.
\cite{Kahabka95b} proposed the 1E0035.4--7230 and Nova Muscae 1983 to be ``supersoft polars'', a subclass of SSSs that the magnetized WD with a strong field of $\sim 10^{7}$--$10^{9}$\,G has synchronized the companion, and the accretion column on the WD surface makes the observed orbital modulation. 
V4743 Sgr, V959 Mon, and \NH\ have different spin and orbital periods, and a polar scenario is highly unlikely though the system can still be an intermediate polar with a magnetic field strength of $10^{5}$--$10^{6}$\,G \citep{Hillman2019,Drake2021,Orio2022}. 
Furthermore, the X-ray folded light curve of \NH\ (Fig.~\ref{FLC}) probably shows a double-hump profile, which is different from the rotational profiles seen in 1E0035.4--7230 and Nova Muscae 1983 \citep{Kahabka96}.

The $\sim$0.153-day X-ray modulation could originate from occultation by the companion.
Besides, it might be explained by occultation by a non-disrupted or re-formed accretion disk seen in high inclination, e.g., V959~Mon \citep{Peretz2016}.
If occultation is the origin, the inclination of the system will be comparable to that of Nova Muscae 1983 (50\degr\--70\degr; \citealt{DS94}) and V959~Mon (60\degr--80\degr; \citealt{Shore2013}). 
In addition, we are aware of V1494~Aql and V5116 Sgr, which have several properties very similar to \NH, e.g., fast novae, similar orbital periods of $\sim$3--4\,hr, double-hump orbital profiles detected in the optical band \citep{BG2003,Kato2004,DRL2008}. 
It will be intriguing to have a deep investigation on the correlation of these three SSSs in the future.

Including \NH, several novae have been detected in GeV $\gamma$-rays before the SSS phase \citep{Gordon2021}, and \NH\ is the only nova system that reveals detectable $\gamma$-ray emission in just $< 1$ day, which could be the shortest duration known for a $\gamma$-ray nova.  
Because the significance of the $\gamma$-ray counterpart is not high (i.e., $\sim$5.8$\sigma$) and the $\gamma$-ray duration is short (i.e., less than a day), the limited photon statistic does not allow us to directly measure the spin and orbital periodicities.
We therefore performed a phase-resolved likelihood analysis to examine the possible spin or orbital $\gamma$-ray modulations, and only the orbital folded light curve provides a marginal indication (Fig~\ref{fig:period_lc}).
If we rely on such a single broad Gaussian structure as the $\gamma$-ray orbital modulation, it is quite distinctive compared with the double-hump structure resolved in the optical band, indicating a different origin. 
The $\gamma$-ray emission is thought to be correlated with the shocks internal to the nova ejecta as the late/fast and early/slow nova winds collide \citep{Metzger2015,Steinberg2020}. It is possible that the fast wind was partially blocked by the companion to create inhomogeneous shock emission, and a quasi-periodic modulated $\gamma$-ray feature could be formed.
To clarify the $\gamma$-ray emitting nature, it is worth checking whether the spin or orbital modulation can be detected in other novae (e.g., ASASSN-16ma; \citealt{Li2022}). 

\section*{Acknowledgments}

This work made use of archival data provided by the LAT data server of FSSC.
This work is supported by the National Science and Technology Council (NSTC) through grant Nos. 110-2811-M-006-012 and 110-2112-M-006-006-MY3.
C.-P.~H. also acknowledges support from the NSTC in Taiwan through grant No. 109-2112-M-018-009-MY3.
J.~T. acknowledges support from the National Key Research and Development Program of China (grant No. 2020YFC2201400) and the National Natural Science Foundation of China (grant No. 12173014). 
K.L.~L. is supported by the NSTC of Taiwan through grant No. 111-2636-M-006-024, and he is also a Yushan Young Fellow supported by the Ministry of Education of the Republic of China (Taiwan).

\section*{DATA AVAILABILITY}

The \emph{NICER} and \emph{Fermi} observations used in this paper are publicly available at the Data Server. \\
\emph{NICER}: \href{https://heasarc.gsfc.nasa.gov/db-perl/W3Browse/w3table.pl?tablehead=name\%3Dnicermastr\&Action=More+Options}{https://heasarc.gsfc.nasa.gov/db-perl/W3Browse/}\\
\href{https://heasarc.gsfc.nasa.gov/db-perl/W3Browse/w3table.pl?tablehead=name\%3Dnicermastr\&Action=More+Options}{w3table.pl$?$tablehead=name\%3Dnicermastr\&Action=More+Options} \\
\emph{Fermi}-LAT: \href{https://fermi.gsfc.nasa.gov/ssc/data/access/}{https://fermi.gsfc.nasa.gov/ssc/data/access/}


\end{document}